# Nanoscale Transport of Surface Excitons at the Interface between ZnO and a Molecular Monolayer

Sebastian Friede[†*], Sergei Kuehn [†], Sergey Sadofev [‡], Sylke Blumstengel [‡],

Fritz Henneberger [‡] and Thomas Elsaesser [†]

*[†] Max-Born-Institut für Nichtlineare Optik und Kurzzeitspektroskopie, D-12489 Berlin,*

*[‡] Institut für Physik, Humboldt-Universität zu Berlin, D-12489 Berlin, Germany*

Excitons play a key role for the optoelectronic properties of hybrid systems. We apply near-field scanning optical microscopy (NSOM) with a 100-nm spatial resolution to study the photoluminescence of surface excitons (SX) in a 20 nm thick ZnO film capped with a monolayer of stearic acid molecules. Emission from SX, donor-bound (DX), and – at sample temperatures T>20 K – free (FX) excitons is separated in steady-state and time-resolved photoluminescence spectra. The 4 meV broad smooth envelope of SX emission at T<10 K points to an inhomogeneous distribution of SX transition energies and spectral diffusion caused by diffusive SX transport on a 50 nm scale with a diffusion coefficient of $D_{SX}$(T<10K)=0.30 cm²/s.

PACS number(s): 68.37.Uv, 73.20.At, 78.20.-e, 78.47.D-, 78.66.Hf

*Corresponding author: friede@mbi-berlin.de, Fax +49 30 6392 1489

Most semiconductors with a direct bandgap $E_g$ display strong excitonic features in their optical absorption and photoluminescence (PL) spectra close to $E_g$. In general, both free Wannier excitons and localized excitons contribute to the optical spectra. Their lineshapes reflect the dephasing of optical polarizations by radiative recombination and/or scattering processes as well as a distribution of exciton binding energies due to structural disorder in the semiconductor [1]. Localized excitons have been applied as local probes of disorder potentials at the surface of or the interface between semiconductor layers [2-4]. In GaAs/AlGaAs quantum wells of high structural quality, the broad spatially averaged PL spectra break up into narrow lines which originate from excitons localized at different sites and display a spectral width determined by the radiative recombination time.

Different types of excitons have been distinguished in the prototype wide gap semiconductor ZnO: free excitons (FX) with a binding energy of ~60 meV, excitons bound to different donor and acceptor atoms (DX), and surface excitons (SX) [5-12]. SX PL has been detected from nanostructures at low temperature [9-11], and from ZnO bulk single crystals or epitaxial layers [8,12]. Recently, we have observed a distinct SX PL band from a ZnO thin film covered with a molecular monolayer [12]. Based on time-resolved SX PL measurements, a model involving a population exchange between less localized and strongly localized states in a heterogeneous potential has been proposed [9]. Based on the enhancement of SX PL by the presence of inert polymers, another model considers the SX a charge-separated state of the FX in the inhomogeneous field at the semiconductor surface [10]. A similar picture has been proposed in the context of photoelectron emission studies on ZnO(10-10) [13].

Little is known about the spatial distribution and localization of the SX and its intrinsic emission lineshape. The emission linewidth of ~4 meV observed even at T<10 K [8,12] could result from homogeneous broadening by dephasing processes or an inhomogeneous distribution of emission energies in combination with spectral diffusion processes. The latter may involve diffusive SX motions between a multitude of sites. In this Letter, we study the SX and DX PL from a thin epitaxial ZnO layer covered with a molecular monolayer of stearic acid in a wide temperature range and with an unprecedented 100 nm spatial resolution. The SX near-field PL shows a smooth 4 meV broad spectrum even at T<10 K, pointing to an inhomogeneous distribution of SX states, in energy partially overlapping with DX states. In contrast, spectral narrowing is observed for the DX emission from particular sites. Time-resolved NSOM PL measurements provide direct evidence for diffusive SX motion in the ZnO layer that occurs with a diffusion coefficient of $D_{SX}$=0.30 cm²/s at T<10 K and results in spectral diffusion of the PL.

Our thin-film sample consists of a pseudomorphic ZnO/ZnMgO layer sequence (Fig. 1a,b) grown on a ZnO single crystal by molecular beam epitaxy. The 20 nm thick ZnO top layer with an oxygen terminated (000-1) surface is followed by 350 nm thick $Zn_{0.88}Mg_{0.12}O$, a 10 nm thick $Zn_{0.85}Mg_{0.15}O$ barrier layer, and a 150 nm thick ZnO buffer on the substrate. On the top ZnO layer, a monomolecular stearic acid film is attached by the Langmuir-Blodgett (LB) technique.

Steady-state PL spectroscopy was performed with continuous-wave (cw) excitation at 3.81 eV by a HeCd laser and a detection system consisting of a grating spectrometer with a 0.3 meV spectral resolution and a cooled CCD camera. Frequency-doubled 150 fs pulses from a modelocked Ti:sapphire oscillator centered at 3.54 eV served for excitation in time-resolved PL experiments. Time-resolved detection was based on time correlated single photon counting with a spectral bandwidth of 1.2 meV and an instrument response of 65 ps (FWHM).

High spatial resolution is achieved with a home-built variable-temperature Nearfield Scanning Optical Microscope (NSOM) operated in excitation/collection mode (Fig. 1a): The PL is collected through an uncoated fused silica fiber tip used also for excitation. At a tip-sample distance of 10 - 20 nm stabilized by a shear-force feedback mechanism, a spatial resolution of 100 nm is achieved. We refer to this arrangement as near-field. In the far-field arrangement with a tip-sample distance of 10 μm, the resolution decreases to >1 μm. Near- and far-field data were collected with the same excitation *density* by adjusting the pump intensity accordingly.

The transport measurement relies on the fact that lateral exciton transport in the ZnO top layer results in a decrease of exciton density within the excitation spot and, thus, accelerates the time-resolved PL decay. The exciton density n(r,t) is described by a rate equation with a diffusional component and a recombination term (Figs. 1d-f). The nearfield PL signal is given by I(t)~∫ n(r,t)·$\eta_{coll}$(r)·r·dr, where $\eta_{coll}$(r) is the collection profile. Using an analytical solution for a Gaussian initial distribution n(r, t=0)~$e^{-r^2/\sigma^2}$ with σ the half width at 1/e [14], and assuming that $\eta_{coll}$(r) ~ n(r, t=0), we arrive at:

$$I(t) = A \frac{e^{-t/\tau}}{2 \cdot D \cdot t + \sigma^2} \qquad (1)$$

where D is the diffusion constant, τ the recombination time and A a constant. For $\sigma^2 \gg 2 \cdot D \cdot \tau$, the denominator becomes time-independent and I(t)~$e^{-t/\tau}$. This condition met in the far-field allows for determining τ. In the near-field, $\sigma^2 \leq 2 \cdot D \cdot \tau$ and the additional change in I(t) yields the diffusion constant D.

For cw excitation at 3.81 eV as applied in the NSOM PL measurements at a sample temperature T<10 K, about 30 percent of the excitation light is absorbed in the 20 nm thick ZnO top layer and the remaining 70 percent in the ZnMgO barrier. The time-averaged excitation density $N_{ex}$ in the ZnO top layer is below one exciton within the illuminated area of ~200×200 nm², i.e., below $N_{ex}$=2.5×10$^9$ cm$^{-2}$. Figure 2a shows a false color image of 20 spectra taken along an NSOM scan line of 1 µm in 50 nm steps. The PL spectra display two prominent bands identified as the DX band at 3.361-3.363 eV and the SX band at 3.365-3.369 eV [12]. The spectra reveal a spatial inhomogeneity in the DX signal amplitude and energy. At x=0.65 µm, a sharp DX line at an energy of 3.3615 eV emerges from the broad background. This feature is limited to the spatial and spectral resolution of the NSOM-setup (Fig. 2b). In contrast, the PL spectra of the SX always remain broad and featureless showing a width of ~4 meV (FWHM, Fig. 2c). DX PL from the substrate is absent when exciting the sample from the top as the excitation light at 3.81 eV is fully absorbed in the layers above.

A set of T-dependent far-field PL spectra is shown in Fig. 2d. Between T<10 K and T=20 K, the SX intensity rises relative to the DX intensity. For T>20 K, both DX and SX intensities decrease and the FX PL band develops at high emission energies. Such changes reflect the thermal population distribution of the different exciton species which is established within the lifetime of the excitonic states. The inset of Fig. 2d shows the temperature dependent peak positions of DX and SX emission which both decrease linearly with temperature at a rate of approximately -32 µeV/K, reflecting the temperature-induced change of the ZnO bandgap.

Time-resolved PL data were taken with excitation pulses at 3.54 eV which excite the ZnO top layer and are fully absorbed in the non-emitting ZnO buffer, i.e., the recorded emission originates from the 20 nm ZnO top layer exclusively. The excitation density was $N_{ex} \approx 6 \times 10^{10}$ cm$^{-2}$. Fig. 3a shows far-field (blue line) and near-field transients (red line) in a 1.2 meV wide spectral window around the SX emission maximum. The near-field data represent an average over 100 sampling points. The near-field PL-transient exhibits an accelerated PL decay compared to the far-field data, a hallmark of in-plane transport of the SX species. The shape of the SX far-field transient closely resembles a single-exponential with a decay time of $\tau_{SX}$=90 ps.

The corresponding measurement at the DX PL maximum again displays an accelerated decay in the near-field transient (Fig. 3b). The DX far-field transient is mono-exponential over more than three decades with a decay time of $\tau_{DX}$=160 ps. The SX and DX transients at short times are compared in Fig. 3c. The rise of DX PL slightly lags behind the SX and reaches its maximum 45 ps after the SX. A kinetic analysis gives a rise time of ~30 ps which represents the population time of DX states after excitation, in good agreement with literature values for bulk ZnO [15]. When the sample is heated to 50 K, the PL spectrum develops into a single broad band with the three exciton species DX, SX and FX still recognizable as shoulders (cf. Fig. 2b). The far-field transients recorded at the three resonances blend into a single one with identical rise and decay time (Fig. 3d). The latter is $\tau_{50K}$=185±5 ps as derived from a single-exponential fit.

When a sharp tip approaches an emitter closely, a number of other effects could potentially lead to a change in the lifetime. We have experimentally ruled out local sample heating by the room-temperature tip and the Purcell effect. All these effects lead to an increase rather than a decrease of decay time. The change of decay time with the tip-sample distance falls off on a 1 µm length scale, in contrast to ~100 nm as predicted for the Purcell effect [16]. Approaching a blunt tip with an apex of 1 µm did not change the apparent lifetime, excluding an influence of the dielectric tip on the surface potential as a cause for the lifetime change [17]. A dielectric tip may open additional radiative channels in the emitter near-field by scattering, leading to a reduction of the radiative lifetime of up to 7% and a small dielectric artefact [18,19]. The DX radiative rate at T<10 K is on the order of 1 ns [20], resulting in a negligible shortening of the measured decay time $\tau_{DX}$= 160 ps by ~1%.

The spectral sharpening of the steady-state DX emission at particular spatial positions suggests that individual local DX sites are selected in the nearfield. The light collection depth of the NSOM is larger than the 20 nm thickness of the ZnO film and, consequently, DX sites distributed over the full film thickness contribute to the recorded PL. A minimum average distance of emitting DX centers of 200 nm, much larger than the exciton Bohr radius of some 2 nm, is estimated from the density of distinct DX lines in the scan image. On the other hand, the typical defect density in high-quality ZnO of $\sim 10^{17}$ cm$^{-3}$ gives an average distance of binding sites of typically of 20 nm (area density $2 \cdot 10^{11}$ cm$^{-2}$). For the much smaller time-averaged excitation density $N_{ex}=2.5\times 10^9$ cm$^{-2}$, only a minor fraction of localization sites contributes to the observed narrow emission and coupling between such sites can be ruled out safely.

The nearfield SX emission measured at low temperature displays a smooth and spectrally broad PL band. In a picture of predominant homogeneous broadening, the 4 meV spectral width would correspond to an excitonic dephasing time on the order of 300 fs. At low sample temperatures with $kT \leq 1$ meV, lattice excitations of ZnO and molecular vibrations in the monolayer are essentially frozen out and cannot induce such a fast dephasing. Dephasing by scattering processes of photoexcited carriers would affect the much narrower DX lineshape as well and can be ruled out. Thus, a predominant homogeneous broadening is inconsistent with the data.

We consider the spectral width of the SX PL a measure for inhomogeneous broadening, originating from a distribution of SX transition energies. Adsorption of molecules plays a key part in the formation of SX centers in the spatially inhomogeneous surface potential [8,12]. In our sample, the average distance between adsorbate molecules is 0.4 nm (density $5.5 \cdot 10^{14}$ cm$^{-2}$), whereas the distance between oxygen atoms in the (000-1) ZnO surface is 0.32 nm. Such sub-nanometer lengths and the mismatch of the distance between adsorbates and the O-O distance set the shortest length scale of disorder, i.e., roughness of the surface potential, to a scale substantially shorter than the exciton Bohr radius of approximately 2 nm (Fig. 1c). Apart from local charges at the ZnO surface, the ZnO units and the carboxylic anchor groups of the molecular layer both display a dipole moment, resulting in strong local electric fields at the interface. An estimate of the interaction energy of the dipole moment of the molecules ($\mu_{SA}$~1 Debye) with one hexagonal cell of the first ZnO layer ($\mu_{ZnO}$~26 Debye) gives a value of ~5 meV, a lower limit for local variations of the surface potential. The exciton wavefunction averages over about 70 molecular sites and is susceptible to such variations of the disorder potential, resulting in a large manifold of SX states densely spaced in energy.

The nearfield SX lineshape represents a convolution of the energy distribution of SX states with the lineshape of individual SX emitters. For the PL linewidth of an individual SX, a lower limit of approximately 20 μeV is estimated from the PL decay time of 90 ps, much smaller than the 4 meV width of the SX PL. The observation of such narrow emission components, i.e., of a break-up of the broad spectrum into many narrow lines, would require that their energy position remains unchanged within the PL lifetime. This condition is not fulfilled here. Apart from charge fluctuations at the surface which may occur within the ~100 ps PL decay time, lateral SX transport between different SX sites leads to spectral diffusion. Within the PL lifetime, a particular SX explores many different sites with different transition energies, in this way broadening the PL spectrum. We conclude that the time-averaged SX PL lineshape is inhomogeneously broadened and affected by SX diffusion.

This picture is validated by the transport measurements of Figs. 3a. The accelerated SX-PL decay rate is due to SX diffusion out of the excitation/detection region. Using eq. (1) with a recombination lifetime of $\tau_{SX}$=90 ps, an SX diffusion coefficient of $D_{SX}$(T<10K)=0.30 cm²/s is derived. Within the recombination lifetime, an SX travels an average distance of 50 nm, probing many SX sites with different PL transition energies and broadening the time-averaged PL lineshape (Fig. 2b). The diffusion coefficient for the SX at T<10 K is similar to literature values of $D_{FX}$=0.6 cm²/s for FX in bulk ZnO and $D_{QW}$=0.25 cm²/s in a quantum well structure [21].

The DX PL data (Fig. 3b) display an enhanced decay rate under near-field conditions as well. Applying eq. (1), one derives a diffusion coefficient D(T<10K)=0.36 cm²/s which agrees with $D_{SX}$(T<10K) within the experimental accuracy. The steady-state PL spectra of Fig. 2b demonstrate an overlap of the low-energy tail of the SX emission spectrum with the DX PL band. In other words, there exist SX states at energies very close to or even identical with the DX energies. A population exchange between DX and SX states which is favored by the much larger density of SX states in the sample, results in lateral diffusion of bound excitons even at such low emission energies and at a temperature T<10 K.

At T=50 K, the quick exchange between the different excitonic species is facilitated by thermal activation. Since all excitonic states are coupled, there is only a single lifetime of 185 ps and the difference in rise time disappears (Fig. 3d). This finding demonstrates the existence of a thermalized distribution of FX, SX, and DX populations. Under these conditions, the diffusion constants at the individual exciton energies are equal within the experimental error: $D_{SX}$~0.21±0.1 cm²/s, $D_{DX}$~0.20±0.1 cm²/s and $D_{FX}$~0.15-0.08/+0.3 cm²/s.

In conclusion, we have shown that the surface exciton on the ZnO (000-1) surface is weakly localized and displays diffusive in-plane transport. The structureless 4 meV broad photoluminescence band observed with a 100 nm spatial resolution at a sample temperature below 10 K suggests an inhomogeneous distribution of transition energies originating from the short-range disorder at the surface. Mobile surface excitons explore this landscape within the photoluminescence lifetime, resulting in a substantial spectral broadening by spectral diffusion.

The population exchange between donor-bound and surface exciton states results in a concomitant transport of the DX species.

Acknowledgements

This work has been financially supported by the Deutsche Forschungsgemeinschaft in the framework of the Collaborative Research Center SFB 951. The authors thank Monika Tischer for preparing and characterizing the fiber tips.

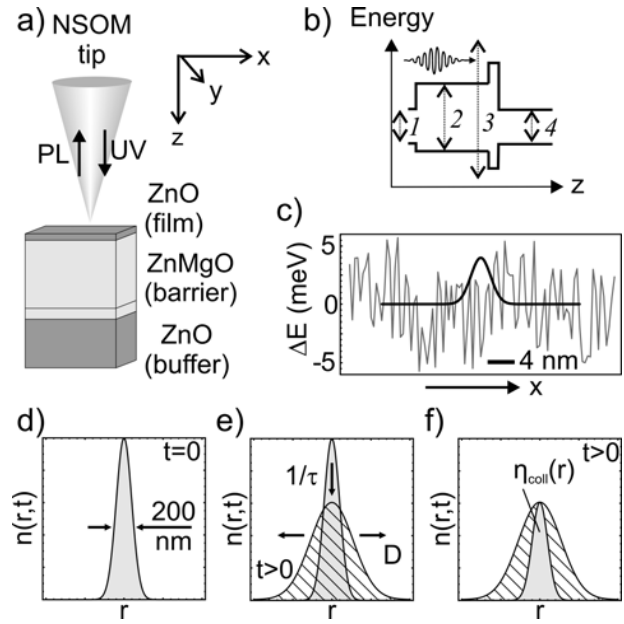

**Figure 1:** (a) Schematic of the experimental setup showing the near-field tip with excitation/collection mode and the sample geometry. **b)** Bandgap diagram and photon energy. *1*: 20 nm ZnO film, 3.37 eV; *2*: 350 nm $Zn_{0.88}Mg_{0.12}O$ barrier, 3.7 eV; *3*: photon energy 3.81 eV cw or 3.54 eV pulsed; *4*: ZnO buffer, 3.37 eV. (c) Sketch of the potential roughness $\Delta E$ and spatial extension of the exciton wavefunction. (d-f) Visualization of (d) the creation, (e) the time evolution by recombination and diffusion, and (f) the detection of the carrier distribution n(r,t) with the NSOM tip.

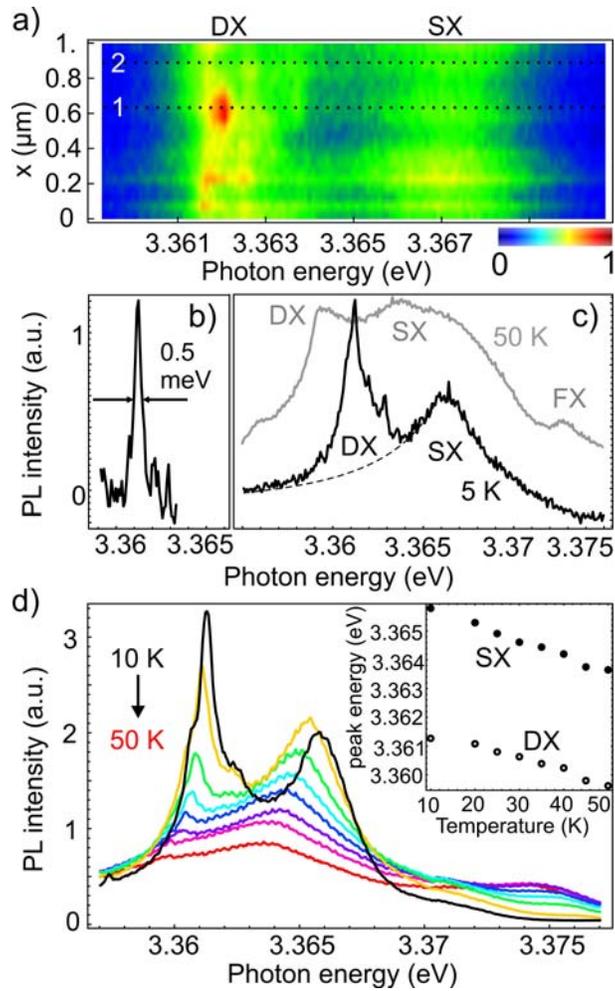

**Figure 2.** (a) Map of 20 PL spectra along a 1 μm scan line at T=5K (excitation at 3.81 eV). (b) Spectrum of an isolated DX peak obtained by subtracting spectrum *2* from spectrum *1*. (c) Black: Near-field spectrum along line (1) in (a). Grey: PL spectrum at 50 K. The thin dashed line is a guide to the eye. (d) Far-field PL spectra for sample temperatures from 10 to 50 K. The black and the orange line gives the PL spectrum for 10 and 20 K, respectively. For measuring the other spectra, the sample temperature was raised in steps of 5 K. Inset: Peak positions of SX and DX emission as a function of sample temperature.

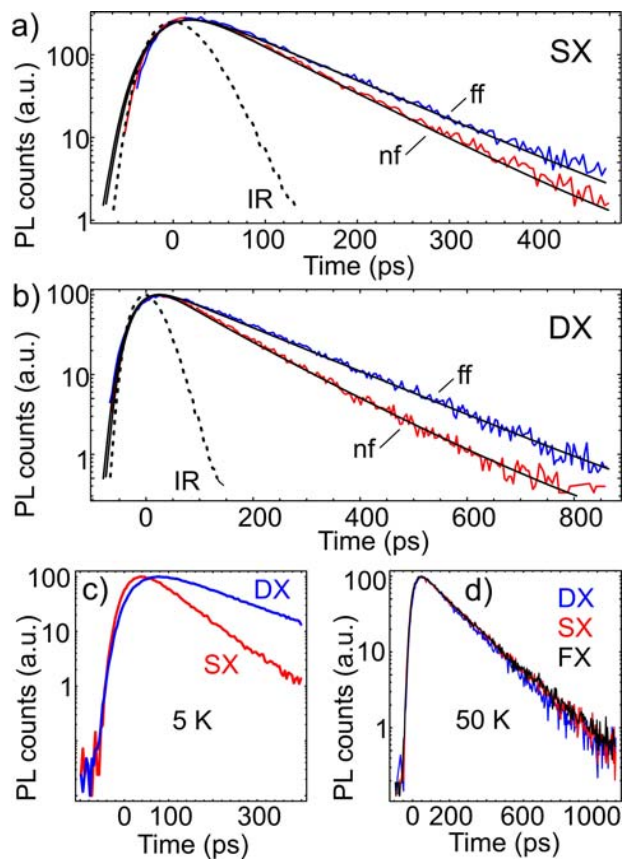

**Figure 3.** Time resolved PL intensity recorded in the near-field (nf) and the far-field (ff) at (a) the SX peak (T=5K) The solid black lines represent numerical fits to the measured kinetics and the dashed line shows the instrument response (IR). (b) Same as (a) at the DX peak. (c) Comparison of the rise times of SX and DX PL in the far-field. **d)** PL transients of the FX, SX and DX at T=50 K.